\documentclass[%
 reprint,
superscriptaddress,
 amsmath,amssymb,
 aps,
]{revtex4-1}

\usepackage{graphicx}
\usepackage{dcolumn}
\usepackage{bm}
\usepackage{verbatim}
\usepackage{color}
\usepackage{gensymb}
\usepackage[version=3]{mhchem}
\usepackage[usenames,dvipsnames]{xcolor}
\definecolor{dark-green}{HTML}{006400}
\usepackage{threeparttable}
\usepackage{adjustbox}
\usepackage{footnote}
\usepackage{float}
\usepackage{hyperref}
\usepackage{scrextend}
\usepackage{tabularx}
\usepackage{chemmacros}
\usepackage{lineno}  %

\newcommand{\lro}{Li$_2$RuO$_3$}
\newcommand{\dlro}{Li$_{2-x}$RuO$_3$}

\newcommand{\ruo}{RuO$_2$}
\newcommand{\td}{$T_{\text{d}}$}
\newcommand{\ptwo}{$P2_1$/\textit{m}}
\newcommand{\ctwo}{$C2/m$}
\newcommand{\io}{I$_2$}
\newcommand{\rufo}{Ru$^{4+}$}
\newcommand{\rufi}{Ru$^{5+}$}

\begin{document}

\preprint{APS/123-QED}

\title{Effect of delithiation on the dimer transition of the \\honeycomb-lattice ruthenate Li$_{2-x}$RuO$_3$}

\author{Marco-Polo Jimenez-Segura}
\affiliation{Department of Physics, Graduate School of Science, Kyoto University, Kyoto 606-8502, Japan}
\author{Atsutoshi Ikeda}%
\affiliation{Department of Physics, Graduate School of Science, Kyoto University, Kyoto 606-8502, Japan}
\author{Simon A. J. Kimber}
\affiliation{European Synchrotron Radiation Facility (ESRF), 6 rue Jules Horowitz, BP 220, 38043 Grenoble Cedex 9, France}
\author{Carlotta~Giacobbe}
\affiliation{European Synchrotron Radiation Facility (ESRF), 6 rue Jules Horowitz, BP 220, 38043 Grenoble Cedex 9, France}
\author{Shingo Yonezawa}%
\affiliation{Department of Physics, Graduate School of Science, Kyoto University, Kyoto 606-8502, Japan}
\author{Yoshiteru Maeno}
\affiliation{Department of Physics, Graduate School of Science, Kyoto University, Kyoto 606-8502, Japan}%

\date{\today}

\begin{abstract}
The honeycomb-lattice ruthenate \lro\ is made heavily Li-deficient by chemical oxidation by iodine. The delithiation triggers a different phase \dlro , the ``D-phase'', with superlattice. For the first time we disclose the magnetic and structural properties of the D-phase in the dimer-solid state. The low temperature magnetic susceptibility and the bond lengths indicate a bonding configuration consisting of both \rufo -\rufo\ and \rufi -\rufi\ dimers.
\end{abstract}

\pacs{Valid PACS appear here}
\maketitle
\section{\label{sec:Introduction}Introduction}
Honeycomb-lattice iridates of the type $A_2$IrO$_3$ (\textit{A}=Li, Na) with the effective angular momentum $J_{\text{eff}}= 1/2$ due to strong spin-orbit coupling have been under active active experimental investigation \cite{ singh_antiferromagnetic_2010, takayama_hyperhoneycomb_2015} mainly because of their interesting  properties associated with predicted topologically non-trivial states, including the Kitaev spin-liquid state \cite{kitaev_anyons_2006, shitade_quantum_2009}. What is more, topological superconductivity has been proposed based on the Kitaev-Heisenberg model with the spin \textit{S} = 1/2 emerging with hole doping \cite{you_doping_2012, hyart_competition_2012, okamoto_global_2013, scherer_unconventional_2014}.

Another interesting compound with the honeycomb structure is \lro . One of the differences from $A_2$IrO$_3$ is that \lro\ has nominally $S=1$, as expected for the low-spin state of Ru$^{4+}$(4$d^4$). Interestingly, \lro\ exhibits dimerization of Ru-Ru ions at \td\ $\approx 540$~K, accompanied by a sharp decrease of magnetization below \td\ \cite{miura_new-type_2007}. More recently, it is found that the disorder in the Ru-Ru dimer configuration sensitively affects the magnetic behavior \cite{jimenez-segura_effect_2016}. In addition, the dimer transition has been revealed to be of the first-order type \cite{terasaki_ruthenium_2015}, especially in samples with more coherent dimer configuratuion with \td ~$\approx 550$~K, \cite{jimenez-segura_effect_2016}.

It has been also demonstrated that the dimer transition is not an ordinary Peierls transition. In a recent study based on a combination of high-energy X-ray diffraction (XRD), pair distribution function (PDF) analysis,  and density functional theory (DFT) calculations, it has been found that the dimers dynamically survives even above \td\ \cite{kimber_valence_2014}. Thus, the transition can be regarded as the change from a static ``dimer-solid'' state at low temperatures to  a dynamic ``dimer-liquid'' state above \td ~\cite{kimber_valence_2014}.

The potential utility of Li$_2$RuO$_3$ as a material for batteries has motivated investigation on its electrochemical properties as well \cite{sarkar_lithium_2014}. 
The delithiated series Li$_{2-x}$RuO$_3$ has been synthesized from Li$_{2}$RuO$_3$ by electrochemical deintercalation of lithium \cite{kobayashi_structure_1995,taminato_mechanistic_2014,mori_relationship_2015} and by chemical oxidation by I$_2$ \cite{kobayashi_physical_1996}. 
Moreover, it has been revealed that a part of \rufo\ changes to \rufi\ by delithiation based on a sequence of experiments \cite{sathiya_reversible_2013, sathiya_high_2013, li_understanding_2016}.
  
However, concerning the physical properties of \dlro , only magnetization has been reported in the limited temperature range between 83 and 293~K \cite{kobayashi_physical_1996}. In particular, the relation between the dimer transition and delithiation has not been reported. 

The effect of Li deficiency on the dimer transition is of primary interest for the following reasons. Firstly, the hole doping by delithiation could lead to the spin $S=1/2$ state and thus to various topological phases including exotic superconductivity. Secondly, the dimer formation may be substantially changed by Li deficiency or hole doping, leading to a possible new dimer state related to dimer-solid and dimer-liquid states.

In this work, we report properties of heavily delithiated \dlro\ obtained by chemical oxidation. We confirm a crystallographic phase distinct from the pristine phase. This phase, emerging by heavy delithiation, shall be called the ``D-phase''. We compare structural and magnetic properties of the D-phase with those of the stoichiometric ``S-phase''. In particular, we find in the D-phase a new dimer-solid state with a different electronic configuration from that in the S-phase.

\section{\label{sec:EXPERIMENT}EXPERIMENT}
Pristine Li$_2$RuO$_3$ samples were prepared from Li$_2$CO$_3$ (Aldrich, 99.997\%) and RuO$_2$ (Rare metallic, 99.9\%) by means of solid-state reaction. After the starting powders were dried, stoichiometric quantities were mixed and ground for 1~h in a conventional mortar. We added acetone to improve the homogeneity of the powder \cite{jimenez-segura_effect_2016}. The powder was pelletized and heated at 1000$^\text{o}$C for 24~h in the first step. Next, the pellet was re-ground in acetone for 1~h, pelletized, and heated at 900$^\text{o}$C for 48~h followed by natural cooling. This choice of the synthesis proccedure is based on the previous study \cite{jimenez-segura_effect_2016}. Purity, as well as the coherence of the  dimer configuration were verified by XRD and magnetic susceptibility measurements as described below.

The delithiation was performed based on the reaction:
\begin{equation}
\text{Li}_2\text{RuO}_3 + \frac{y}{2} \text{I}_2  \xrightarrow[]{\text{CH}_3\text{CN}}  \text{Li}_{2-x}\text{RuO}_3 + x \text{LiI} + \frac{y-x}{2} \text{I}_2
 \label{ch:delithiation}
\end{equation}

I$_2$ (Wako, 99.9\%) was dissolved in acetonitrile (CH$_3$CN) with the concentration of 0.3384 mol/L. After precise measurement of its mass, the powder of Li$_2$RuO$_3$ was soaked in the iodine solution for 3 days at room temperature. 
Then the \dlro\ powder was washed with clean acetonitrile. Since XRD analysis indicated that stronger delithiation was necessary, after another measurement of the mass, the powder was soaked in a new solution of I$_2$ 0.3683 mol/L for 18~h under stirring at 560~rpm. The containers were covered with aluminum foil in order to avoid the conversion of I$^{-}$ into I$_2$ triggered by light. The Li$_{2-x}$RuO$_3$ powder was taken out, and rinsed with clean acetonitrile until the color of the acetonitrile became completely transparent. A less-delithiated sample of \dlro\ was prepared in a solution of \io\ 0.18457~mol/L. In this case, a pellet of \lro\ was soaked for 3 days. We stored samples in vacuum although there is no noticeable decomposition at room temperature in air for both pristine and delithiaed samples.

In order to evaluate the value of \textit{x}, the remaining quantity of I$_2$ in acetonitrile was measured through titration \cite{daniel_c._harris_quantitative_2006, patnaik_deans_2004} with a standard solution of Na$_2$S$_2$O$_3$ (Wako, 0.05 mol/L for volumetric analysis) at 19\celsius\ based on the reaction: 
\begin{equation}
\text{I}_2 + 2 \text{S}_2\text{O}_3^{2-} \rightarrow  2\text{I}^{-} + \text{S}_4\text{O}_6^{2-} \label{ch:titration}
\end{equation}
The glassware used for the titration was calibrated at 19\celsius\ with acetonitrile. The completion of this reaction is monitored by the color of starch added in the solution. We find that the total lithium extracted is $x \approx 0.73$ and 0.34 for the two delithiated samples presented in this paper.
 
We also performed the inductively-coupled plasma optical emission spectroscopy (ICP-OES) analysis using a commercial apparatus (Seiko Instruments, SPS 6100). Quantitative evaluation of \textit{x} by ICP-OES was not successful because \dlro\ cannot be dissolved completely into standard acids (such as HCl), due to production of insoluble \ruo\ in the acid.

The laboratory XRD measurements were carried out with a commercial diffractometer (Bruker, D8 Advance) using the Cu$K \alpha$ radiation ($\lambda=1.54184$~\AA , $E=8.041$~keV) equipped with a one-dimensional array of detectors and a nickel monochromator. 
High-energy XRD measurements at room temperature were performed at the beam line ID22 of the European Synchrotron Radiation Facility (ESRF). X-ray beam of the energy $30.993~\text{keV}~(\lambda=0.40003$~\AA) was used.
High-energy XRD measurements at various elevated temperatures were performed at the beam line ID11 at ESRF. A double Laue monochromator was used to select X-ray of the energy of 87.5~keV ($\lambda=0.14169$~\AA ). The beam was focused to ca.~100 mm using refractive lenses. The scattered X-ray was detected using a CCD camera (FReLoN). Temperature control was achieved using a hot-air blower. Data were continuously collected upon heating to 723~K at 2.5~K/min, and upon cooling to room temperature at the same rate. Typical temperature resolution was 0.2~K/pattern. We find a noticeable difference between the actual sample temperature and the thermometer temperature. Thus, a lineal correction to the thermometer temperature was made, so that \td\ of \lro\ recorded in the XRD measurement matches with the \td\ of the magnetization measurements.

The magnetization measurements were performed using a commercial superconducting quantum interference device magnetometer (Quantum Design, MPMS). The magnetization at high temperatures (300 to 700~K) were measured using the oven option for MPMS. We used quartz tubes as sample holders for the high temperature measurements. The sample tubes were sealed with ceramic bond (Resbond, 907GF) in order to prevent any gas released from the sample to damage the inside of the oven sample space. We checked the thermometer calibration of the oven by measuring the ferromagnetic transition ($T_\text{C}$= 627.2~K) of Ni (Rare metallic, 99.99\%) at several fields \citep{kouvel_detailed_1964}. The calibration error in $T_{\text{C}}$ of Ni is less than 0.2\%. Measurements from 1.8 to 300~K were performed with the ordinary setup of MPMS in zero-field-cooled (ZFC) and field-cooled (FC) sequences.

\section{\label{sec:RESULTS AND DISCUSSION}RESULTS AND DISCUSSION}

\subsection{\label{sec:Crystal Structure}Structure}

\begin{figure}
\includegraphics[width=8.8cm]{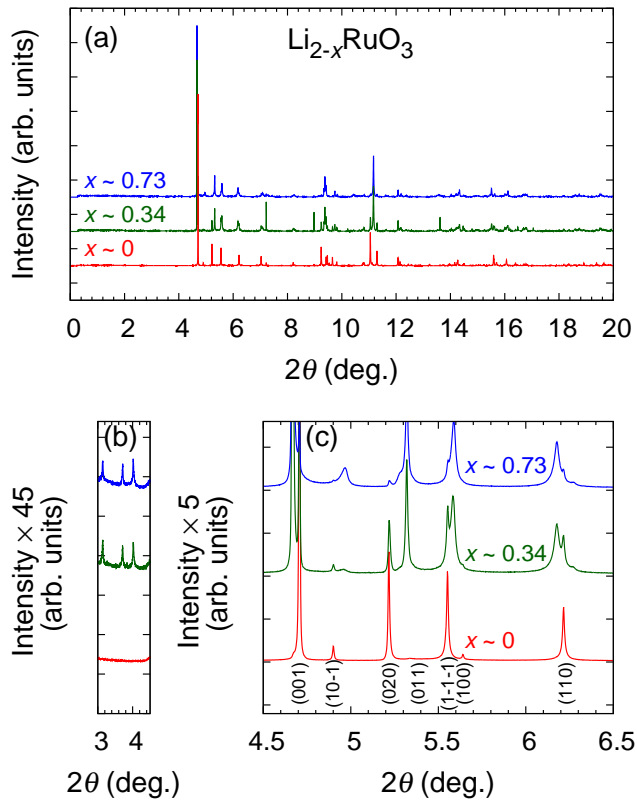}
\caption{\label{fig:XRD_HE} Temperature variations of the high-energy powder XRD spectra at room temperature of \dlro\ samples with $x=0$, 0.34, and 0.73, taken with X-ray of $30.993~\text{keV}~(\lambda=0.40003739(180)$~\AA) at the beam line ID22 of the ESRF. Panels (a), (b), and (c) present data in the ranges $0^\circ< 2\theta < 20^\circ$, $3^\circ< 2\theta < 4.5^\circ$, and $4.5^\circ< 2\theta < 6.5^\circ$, respectively. The peak indices of the stoichiometric \lro\ are shown in (c) based on the room-temperature crystal structure reported by Miura \textit{et al.} \citep{miura_new-type_2007}. The $(10\bar{1})$ peak is characteristic of the dimer-solid state \citep{miura_new-type_2007, kimber_valence_2014}}. 
\end{figure}

\begin{figure}
\includegraphics[width=8.8cm]{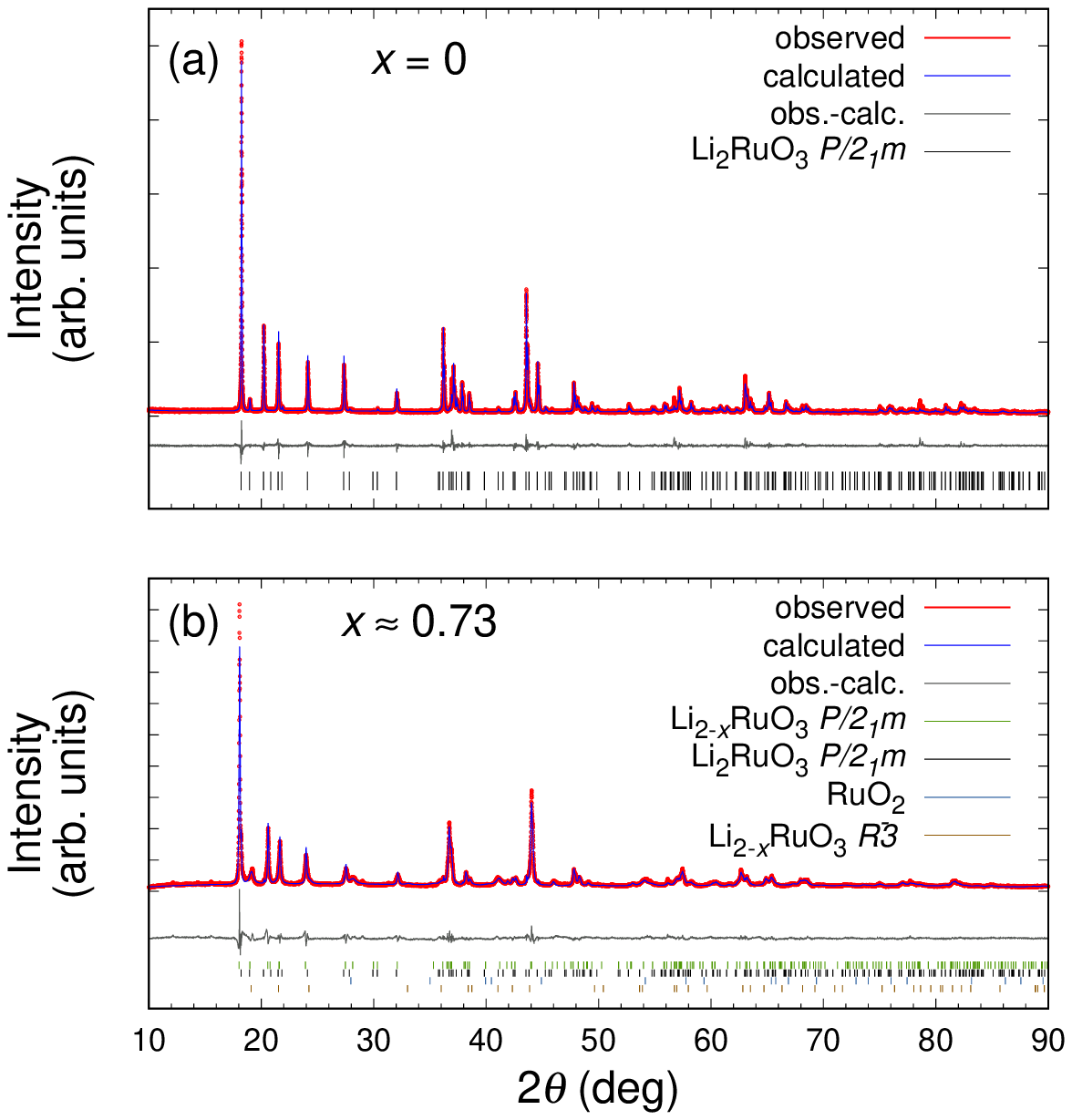}
\caption{\label{fig:Rietveld} (color online) Rietveld refinement of the Cu XRD ($\lambda=1.54184$~\AA , $E=8.041$~keV) of the samples (a) \lro\ with $x=0$ and (b) \dlro\ with $x=0.73$.  In both cases SG \ptwo\ was used. Parameters of the fitting are shown in Table~\ref{tb:celpar}. The structure factor used in both cases is the one of \rufo because the structure factors of \rufi\ are not available in our software. The measurement time for $x=0.73$ was approximately twice longer than that for $x=0$.}
\end{figure}

Figure~\ref{fig:XRD_HE} shows the high-energy powder XRD spectra for Li$_{2-x}$RuO$_3$ samples with various values of \textit{x} measured at room temperature. This provides high-resolution spectra containing clear superlattice peaks at low angles. Figure~\ref{fig:Rietveld} shows Cu$K \alpha $ XRD with Rietveld fitting.
The pristine ($x=0$) sample exhibits the patterns expected for the pure \lro . The spectrum can be well fitted with the space group (SG) \ptwo\ [No. 11, Fig~\ref{fig:Rietveld}(a)] as reported previously \cite{miura_new-type_2007, kimber_valence_2014, lei_structural_2014, terasaki_ruthenium_2015, jimenez-segura_effect_2016}.

The delithiated samples exhibit patterns with substantial differences compared with the $x=0$ sample as described below.
The $x\approx 0.73$ sample, exhibiting almost single-phase behavior, has the (\textit{h}00) and (00\textit{l}) peaks shifted to lower angles and the (0\textit{k}0) peaks shifted to higher angles (Fig.\ref{fig:XRD_HE}(c)), indicating increase in the values of \textit{a} and \textit{c} and decrease in the value of \textit{b} [Fig.\ref{fig:MTlow}(c)]. Nevertheless, the spectrum of this delithiathed sample can be well fitted also with the SG \ptwo\ [Fig~\ref{fig:Rietveld}(b)]. We designate the crystallographic phase observed in the $x\approx 0.73$ sample as the D-phase.
In addition to the main peaks associated with the D-phase, we can identify minor phases such as the S-phase \lro , \ruo , and yet another delithiated phase \dlro\ with $x \approx 1.1$ (SG $R\bar{3}$, No. 148) \cite{kobayashi_structure_1995}.

For the sample with $x\approx 0.34$, we find that the XRD peaks such as (001) and (020) are clearly split into two peaks [Fig.\ref{fig:XRD_HE}(c)]: one group similar to those of the $x=0$ sample and the other similar to those of the $x\approx 0.73$ sample. This fact indicates that the crystal-structure change due to delithiation is not continuous: there are two crystallographically distinct phases with $x=0$ and $x>0$ and they coexist in similar ratio in the $x \approx 0.34$ sample.

In addition to the shifts in the main peaks, three small peaks but sharp peaks are observed in the delithiated samples at 
$2\theta = 3.140^\circ,~ 3.714^\circ,~ \text{and}~ 4.017^\circ $ [\textit{d} = 7.30, 6.18, and 5.71 \AA , Fig.\ref{fig:XRD_HE}(b)], indicating the presence of a superlattice structure. Although these small peaks have not been reported in pristine or delithiated \lro , a superlattice in Li$_{2-x}$Ru$_{1 - y}M_y$O$_3$  ($M=$Mn, Sn) has been recently found as additional spots in electron diffraction \cite{sathiya_high_2013, sathiya_reversible_2013}. In these studies, the structure has approximately been described with the SG \ctwo\ (neglecting the presence of the superlattice and Ru-Ru dimers even for small values of \textit{y}). It has been suggested that the origin of the superlattice is the distortion in the oxygen positions \cite{sathiya_reversible_2013, li_understanding_2016}. Another possibility is an ordering of Ru valency (charge order) as we discuss latter. 
Since the SG in \dlro\ with the superlattice taken into account is not trivial, we hereafter consider the SG \ptwo\ to evaluate the cell parameters and the Ru ion positions. 

\begin{table}
\setlength\extrarowheight{1.1pt}
\caption{\label{tb:celpar}
Cell parameters of \lro\ (S-phase) and \dlro\ ($x \approx 0.73$, D-phase) at room temperature obtained from laboratory XRD spectra. The space group used for both phases is \ptwo .} 
\begin{ruledtabular}
\begin{tabular}{ccccccccc} 
\textit{x}& \textit{a} (\AA)& \textit{b} (\AA)& \textit{c} (\AA)& $\beta$ ($^{\circ}$)& $\sqrt{3} a/b $ &$R_{\text{wp}}$& $R_{\text{e}}$ & GOF\\
\colrule
0 & 4.922 & 8.787 & 5.896 & 124.36 & 0.970 & 10.96 &7.31  &1.50\\
0.73 & 4.937 & 8.630 & 5.898 & 123.54 & 0.991 & 11.35 & 4.37 & 2.60 \\

\end{tabular}
\end{ruledtabular}
\end{table}

\begin{figure}
\includegraphics[width=6cm]{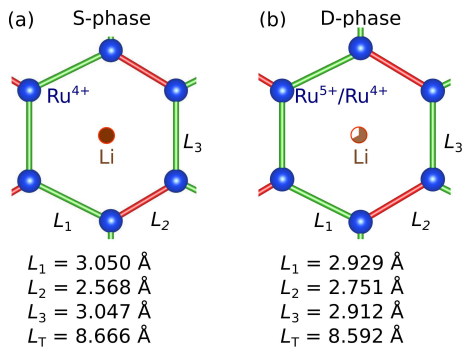}
\caption{\label{fig:bondlength} (color online) Bond lengths $L_1$, $L_2$ and  $L_3$ of the S-phase and D-phase.  The sum of the bond lenghts $L_\text{T}$ is given by $L_1 + L_2 + L_3 $. Blue and brown spheres represent the Ru and Li ions. Red and green bars are the short and long bonds, respectively. The figures are prepared with the program VESTA \cite{momma_vesta_2011}.}
\end{figure}

The cell parameters of the S and D-phases, obtained from the laboratory XRD data, are compared in Table~\ref{tb:celpar}. The parameters \textit{a} and \textit{b} of \dlro\ are longer and shorter than those of \lro , respectively.
Studies on Li$_2$Ru$_{1-y}$Mn$_y$O$_3$ revealed the same trend of change in \textit{a} and \textit{b} when lithium is deintercalated, although the structure of Li$_2$Ru$_{1-y}$Mn$_y$O$_3$ is described with the SG \ctwo ~\cite{sathiya_high_2013}. 
Furthermore, the relation $\sqrt{3}a/b$ of \dlro\ is closer to the unity than that of \lro\ (Table~\ref{tb:celpar}). Thus, on average, the structure of \dlro\ seems to be more symmetric. 
Figure~\ref{fig:bondlength} compares the difference in the bond lengths between the S and D-phases. Reflecting the dimer-solid state in the S-phase, the dimer bond $L_2$ is $\approx$16\% shorter than the long bonds $L_1$ and $L_3$. In contrast, for the D-phase, the difference is only about 6\%. In Sec.~\ref{sec:Electronic}, we will discuss local structures in more detail.

\begin{figure}
\includegraphics[width=8.8cm]{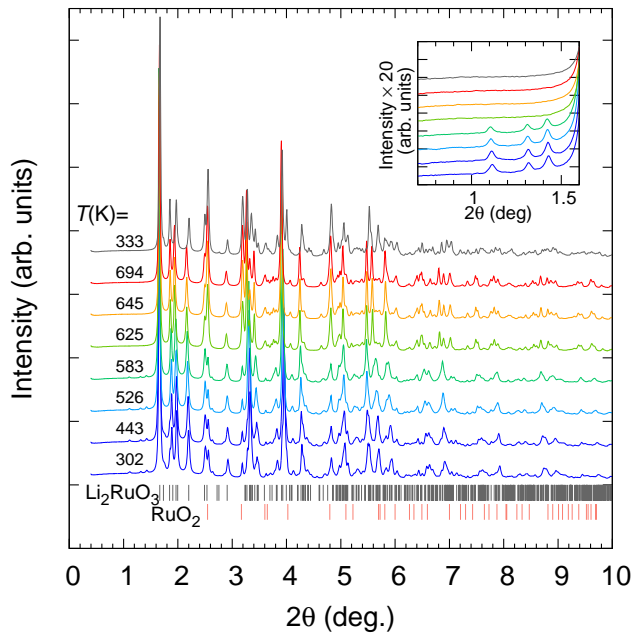}
\caption{\label{fig:XRDlowhi} (color online) Temperature variations of the high-energy XRD spectra for the $x\approx 0.73$ sample with $E=87.5~\text{keV}$ ($\lambda = 0.14169$~\AA) measured at the beam line ID11 of ESRF. Colors of the spectral curves correspond to the temperature region in the magnetization curve of the $x\approx 0.73$ sample in Fig.~\ref{fig:MTlowhi}. The inset shows spectra in the range $0.7^\circ < 2 \theta < 1.5^\circ$. Spectra for 302~K to 664~K are obtained during the heating process, while the spectrum at 333~K was measured after cooling. The vertical lines at the bottom indicate the expected peak positions for \lro\ and \ruo .}
\end{figure}

Figure~\ref{fig:XRDlowhi} shows high-energy XRD patterns ($E=87.5~\text{keV}$, $\lambda = 0.14169$~\AA) of the $x \approx 0.73$ sample (dominated by the D-phase) at selected temperatures on heating (302~K to 694~K) and after subsequent cooling (333~K). While heating, the three superlattice peaks between $2\theta = 1.1^\circ$ and $1.5^\circ$ disappear at $\approx 609$~K. At the same temperature, a sudden increase in the \ruo\ peak intensity, for example the one at $2\theta =3.2^\circ$, is observed. These facts indicate the decomposition of \dlro\ ($ x\approx 0.73 $) into \lro\ and \ruo\ starting from this temperature (see Appendix~\ref{sec:decomposition} for details).

\subsection{\label{sec:Magnetization}Magnetization}

\begin{figure}
\includegraphics[width=8.8cm]{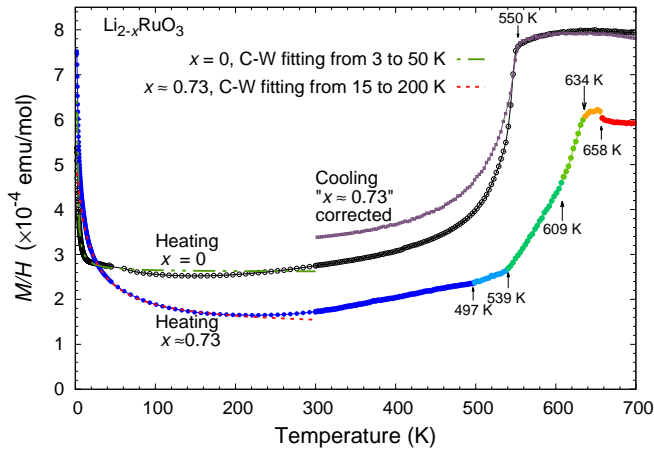}
\caption{\label{fig:MTlowhi} (color online) Magnetic susceptibility vs temperature of the $x=0$ sample (S-phase) and the $x \approx 0.73$ sample (mostly D-phase) from 1.8 to 700~K at \textit{H}~=~10~kOe. Corrections of diamagnetic contributions of ion cores have been made~\cite{bain_diamagnetic_2008}. From 1.8 to 300~K, the magnetization was measured while warming up after ZFC. From 300 to 700~K the magnetization was measured in temperature up sweep, except for the  purple curve which is for the molar susceptibility of \lro\ contained in the $x \approx 0.73$ sample on cooling from 700~K to 300~K (see text). Results of the Curie-Weiss fitting ($\chi = C/(T- \Theta ) + \chi _0$) are also shown with broken curves. Colors of the curve for $x \approx 0.73$ correspond to those in Fig.~\ref{fig:XRDlowhi}.}
\end{figure}

Figure~\ref{fig:MTlowhi} shows the temperature dependence of the  magnetic susceptibility of the $x=0$ and $x\approx 0.73$ samples. Diamagnetic contributions of ion cores have been subtracted \cite{bain_diamagnetic_2008}. The overall shapes of curves for these samples differ in several aspects. 
At low temperatures, both samples exhibit Curie-like behavior; the difference will be discussed later. Both samples exhibit a minimum of magnetic susceptibility but at different temperatures: $\sim 150$~K for the sample with $x=0$ and $\sim 200$~K for the sample with $x \approx 0.73$.  With increasing temperature, the magnetic susceptibility of both samples increases toward the transition from dimer solid to dimer liquid. The change of the magnetic susceptibility in the pristine sample \lro\ is relatively gradual below $\sim 500$~K, followed by a sharp jump characteristic of the first-order transition (Appendix~\ref{sec:mix}).  
In contrast, the sample with $x \approx 0.73$ does not exhibit a gradual change below the transition but exhibits a clearer change in the slope of the $M(T)/H$ curve at $T~\approx 539$~K. Between this temperature and 609~K, no clear change in the averaged crystalline structure was observed and the superlattice peaks are maintained (inset of Fig.~\ref{fig:XRDlowhi}). Thus, the change in the susceptibility seems to be mainly linked to changes in the electronic state, but not to chemical decomposition.

At 609~K, there is another change of slope in the $M(T)/H$ curve (Fig.~\ref{fig:MTlowhi}). As already explained, the decomposition of \dlro\ into \lro\ begins at this temperature (see also Appendix \ref{sec:decomposition}). Therefore, the slope change at 609~K is mainly due to the decomposition. Thus, in the warming run the susceptibility of the $x \approx 0.73$ sample above 609~K represents the sum of those of evolving multiple phases. While warming above 658~K, there is a slight but sharp drop in the susceptibility of the $x \approx 0.73$ sample (dominated by the S-phase at this temperature). Across this temperature, there is no noticeable anomaly in the XRD spectra. Thus, the sharp drop probably reflects a reorganization of lithium or oxygen induced by the vacancies of lithium in the initial  \dlro .

The purple curve in  Fig.~\ref{fig:MTlowhi} is obtained from the susceptibility in the cooling process of the $x \approx 0.73$ sample  after decomposition. Under this condition, the sample is dominated by \lro\ and \ruo\ (see Appendix \ref{sec:decomposition}). The raw data matches with the susceptibility shown by the red curve of the $x \approx 0.73$ sample above $T = 658$~K. The molar susceptibility of \lro\ contained in the decomposed $\approx 0.73$ sample was estimated by using Eq.~\ref{eq:decomposition} and the susceptibility of \ruo\ \cite{fletcher_magnetic_1968}. The result is shown by the purple curve in Fig.~\ref{fig:MTlowhi}, which indeed matches with that of the for the pristine \lro\ above $\sim 550$~K. However, the molar susceptibility shown by the purple curve at room temperature is  $\sim 18\%$ larger than that of the pristine $x=0$ sample. This is attributable to less conherent dimer configuration in \lro\ after decomposition of \dlro ~\cite{jimenez-segura_effect_2016}.

\squeezetable
\begin{table}
\centering
\setlength\extrarowheight{1.1pt}
\caption{Curie-Weiss fitting parameters, expected Curie constant and localized spins.}
\label{tb:C-W}
\begin{tabular}{cccccccc}\hline\hline
\textit{x} & \begin{tabular}[c]{@{}c@{}}Curie \\ const.\\(emu\,K/mol)\end{tabular} & \begin{tabular}[c]{@{}c@{}}$\varTheta$\\(K)\end{tabular} & \begin{tabular}[c]{@{}c@{}}$\chi _0 $\\ (emu/mol) \end{tabular} & \begin{tabular}[c]{@{}c@{}}$\mu _{\text{eff}}$\\ $\mu _{\text{B}}$\end{tabular} & \begin{tabular}[c]{@{}c@{}}Fitting\\ range\\ (K)\end{tabular} & \begin{tabular}[c]{@{}c@{}}Expec. \\ spin \end{tabular} &  \begin{tabular}[c]{@{}c@{}}Loc. \\ spins\\ (\%)\end{tabular} \\
\hline

0  & 0.00039 & 0.96	& 0.0002610	& 0.056	& 3--50 & 1 & 0.039 \\
0.73 & 0.00625 & -16.0 & 0.0001349 & 0.22  & 15--200 & 3/2 & 0.333                                   \\ \hline\hline
\end{tabular}
			
\end{table}

We performed a Curie-Weiss fitting to the susceptibility at low temperatures. The fitting temperature ranges (shown in Table \ref{tb:C-W}) are chosen so that the positive slope in $M(T)/H$ associated with the dimer transition at \td\ does not affect the fittings.
Results of the fitting with the Curie-Weiss law $\chi (T)=\chi _{0}+C/(T-\varTheta)$ are shown with the broken curves in Fig.~\ref{fig:MTlowhi}. From the fittings, we obtain the Curie constant (\textit{C}) and the Weiss temperature ($\varTheta$) as $C = 0.00039$~emu\,K/mol and $\varTheta = 0.96$~K for the $x = 0$ sample, and $C = 0.00625$~emu\,K/mol and $\varTheta = -16$~K  for the $x \approx 0.73$ sample. These results imply that the number of localized spins is rather small even for the $x \approx 0.73$ sample.

In an earlier study, it is proposed that samples with broad magnetic transitions at \td\ are accompanied by a dimer decoherent configuration, namely dimer patterns breaking the long-range ordered configuration \cite{jimenez-segura_effect_2016}. Such dimer decoherent configuration results in nondimerized Ru ions with finite spin. To examine the number of such nondimerized ions, we assume that they exhibit $S=1$ and $S=3/2$ for the S-phase and D-phase, respectively, and that they can be treated as nearly free spins. From the Curie-Weiss fitting described above, the quantity of nondimerized ions is 0.039\% for the $x=0$ sample and 0.33\% for the $x \approx 0.73$ sample.  It is important to note that the obtained number of nondimerized ions in the $x \approx 0.73$ sample is much smaller than the number of vacancies of lithium, or equivalently the number of \rufi\ ions. This fact indicates that most of the Ru ions form dimers even in the D-phase.

\begin{figure}
\includegraphics[width=8.5cm]{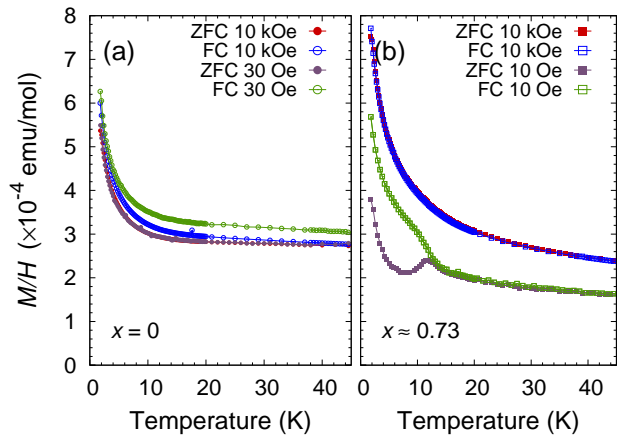}
\caption{\label{fig:MTlow} Magnetic susceptibility vs temperature of (a) \lro\ and (b) \dlro\ ($x \approx 0.73$) from 1.8 to 45~K at different magnetic fields measured in ZFC (closed symbols)  and FC (open symbols) processes.}
\end{figure}

Figures ~\ref{fig:MTlow}(a) and (b) compare the low-temperature susceptibilities for the $x=0$ and $x \approx 0.73$ samples. It is noticeable in Fig.~\ref{fig:MTlow}(b) that the sample with $x \approx 0.73$ at low fields exhibits magnetic hysteresis below 12~K. Since the Weiss temperature $\varTheta$ is negative, this hysteresis is probably due to magnetic ordering with antiferromagnetic interactions. No trace of superconductivity was found by AC susceptibility measurements using an adibatic demagnetization refrigerator \cite{yonezawa_compact_2015} or transport measurements down to 0.1~K. 

\subsection{\label{sec:Electronic} Electronic configuration of dimers}
According to the outcome of the last section, the D-phase is dominated by Ru-Ru dimers. In this section, we discuss the electronic configuration of these dimers and their correlation with the crystalline structure.

We first discuss the probable valency of Ru ions in the D-phase. 
It has been demonstrated by X-ray photoemission spectroscopy, M\"{o}ssbauer spectroscopy, electronic paramagnetic resonance measurements, and DFT calculations, that the deintercalation of lithium from \lro\ leads to the valence change from \rufo\ to \rufi . In higher \textit{x} range than $x=1$, where Ru ions are fully oxidized to pentavalent, it is revealed that further delithiation induces loss of oxygen, presence of peroxides (O$_{2}^{2-}$),  and oxidation of \rufi\ into Ru$^{6+}$ ions \cite{sathiya_reversible_2013, sathiya_high_2013, li_understanding_2016}. Thus, in the sample of the present study with $x \approx 0.73 $, the ratio of the amount of \rufi\ among the total Ru ions would be similar to this \textit{x} value and the others remain \rufo . In other words the valency of the sample can be approximately expressed as \ce{Li_{2-$x$}Ru^{4+}_{1-$x$}Ru^{5+}_{$x$}O3}.

\begin{figure}
\includegraphics[width=6cm]{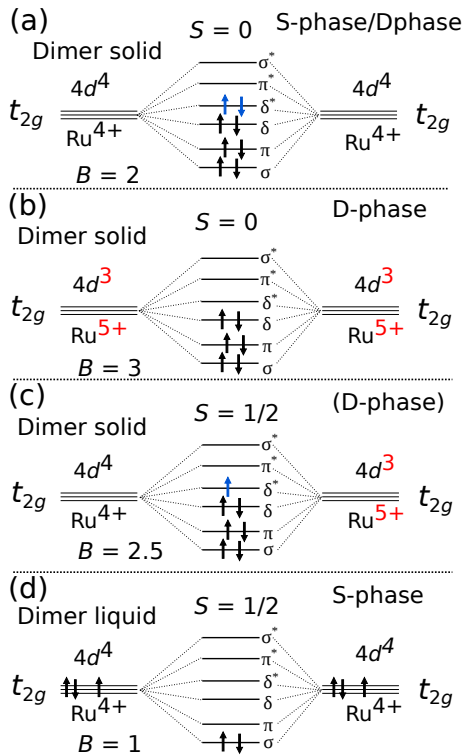}
\caption{\label{fig:MO} Molecular orbital diagrams of possible configurations of the dimers of \lro\ (S-phase) and \dlro\ (D-phase). Panels (a), (b), and (c) depict the molecular orbitals in the dimer-solid state consisting of Ru$^{4+}$--Ru$^{4+}$, Ru$^{5+}$--Ru$^{5+}$, and Ru$^{4+}$--Ru$^{5+}$ dimers, respectively. Configuration in the panel (c) is less likely to occur based on the observed magnetic susceptibility. Panel (d) depicts the molecular orbitals in the dimer-liquid state formed by Ru$^{4+}$--Ru$^{4+}$ ions proposed in Ref.~\cite{kimber_valence_2014}. \textit{B} is the bond order defined by Eq.~(\ref{eq:bon_ord}).}
\end{figure}

For the S-phase, it is expected that the magnetization is governed by the Ru$^{4+}$~(4$d^{4}$) ion. The origin of the decrease of the magnetization below \td\ has been attributed to the formation of dimers with molecular orbitals (MOs) with \textit{S}~$\approx$~0~ \cite{miura_structural_2009}. In contrast, the greater magnetization above  \td\ has been explained by the partial breakdown of MOs. In the dimer-liquid state above \td , the Ru~4\textit{d} orbitals keep the MO formed only by the $\sigma ~\text{bonds}$, leading to dynamic dimers~\cite{kimber_valence_2014}.

Since the Curie-Weiss analysis indicates that most of the Ru ions form dimers with $S=0$ even in \dlro , there are three possible electronic configurations of the dimers in the dimer-solid state as depicted in Fig.~\ref{fig:MO}. The configuration in panel (a) is what already proposed for the S-phase (Refs.~\cite{miura_new-type_2007, miura_structural_2009, kimber_valence_2014}).

We propose that in the D-phase the most probable  configuration for \rufi\ is the dimer formed between \rufi\ and \rufi\ [Fig.~\ref{fig:MO}(b)], since such a dimer has electrons only in the lower-energy bonding states. In contrast, the dimers formed between \rufo\ and \rufo\ ions or \rufo and \rufi\ ions [Fig.~\ref{fig:MO}(a, c)] contain electrons also in higher-energy antibonding states. Actually, the amount of dimers with the configuration \rufi -\rufo\ accompanied by spin $S=1/2$ must be quite small based on the observed small Curie constant.

The situation that most of the Ru ions form isovalent-spinless dimers is actually difficult to be realized if \rufi\ and \rufo\ ions distribute randomly.
Thus, the Ru ions in the D-phase may exhibit charge ordering between \rufo\ and \rufi ions. Such charge ordering may be the origin of the observed superlattice. 

Since \rufo --\rufo\ dimers contain a MO with higher energy than those of \rufi --\rufi\ dimers i.e. $\delta ^*$, it is more likely that the $\delta ^*$ of the  \rufo --\rufo\ dimers breaks at lower temperature than MOs of \rufi --\rufi [Fig.\ref{fig:MO}(b)]. The \td\ where MOs of \lro\ break has been demonstrated to depend on the dimer coherence configuration \cite{jimenez-segura_effect_2016}. Thus, similar to the S-phase, the enhancement of magnetization in the D-phase at $\sim 539$~K (Fig.~\ref{sec:Magnetization}) is probably related to the breaking of MOs of the \rufo --\rufo\ dimers.

In the context of the linear combination of atomic orbitals, the bond lengths \textit{L} of different dimer configurations can be estimated in terms of the quantity called the bond order \textit{B}, which is evaluated as 
\begin{equation} \label{eq:bon_ord}
B = \frac{1}{2}(n_\text{b} - n_\text{a}),
\end{equation}
where $n_\text{b}$ and $n_\text{a}$ are the numbers of electrons in the bonding and antibonding states, respectively. In Fig.~\ref{fig:MO}  the values of \textit{B} of the possible electronic configurations are shown. 
Pauling found an empirical relation between \textit{B} and the bond length $L$  \cite{pauling_atomic_1947, pauling_crystallography_1970}: 
\begin{equation} \label{eq:bon_leng}
L=L_\text{0}-f~\text{log}_{10} (B),
\end{equation}
where the value of \textit{f} depends on the atoms and $L_\text{0}$ is the bond length for $B=1$. Later this relation was derived from the Friedel model \cite{adrian_p._sutton_electronic_1993}. In order to evaluate the constant \textit{f}, we use the value of the bond length of the dimer-liquid state of the S-phase ($L_\text{0} = 2.68$~\AA )~\cite{kimber_valence_2014}, where only the $\sigma $-MO is occupied i.e. $B=1$ [Fig.~\ref{fig:MO}(d)]. We also use the bond lengths of the short bond in the dimer-solid state of the S-phase ($L = 2.58$~\AA ) where the $\sigma,~ \pi,~  \delta ,~\text{and}~ \delta ^{\ast} $ MOs are occupied i.e. $B=2$ [Fig.~\ref{fig:MO}(a)]. From these values, we obtain $f= 0.332$~\AA . By using these values we estimate the length of the \rufi --\rufi\ bond in the dimer-solid state to be $L = 2.52$~\AA .

Since in the dimer-solid state of the $x \approx 0.73$ sample, around 73\% of the dimers are formed by \rufi --\rufi\ ions and the rest by
by \rufo --\rufo ions,  the average length of the short bonds is expected to be 2.56~\AA\ if we assume that the dimers are all located in the $L_2$ bonds. This is 0.04~\AA\ shorter than the dimer entirely formed between   \rufo and \rufo\ ions in the S-phase. On the other hand, the observations for the sum of the bond lengths ($L_\text{T}=L_1 + L_2 + L_3 $) of the D-phase ($L_\text{T}=$8.592~\AA) is 0.074~\AA\ shorter than that of the S-phase ($L_\text{T}=$8.666~\AA ). The excess of shrinkage of $L_\text{T}$ maybe due to the difference in ionic size between \rufo\ ($r=0.62$~\AA ) and \rufi\ ($r= 0.565$~\AA ) \cite{shannon_revised_1976} in the nondimer bonds, although covalency needs to be considered.  
The reduced difference between short and long bonds in the D-phase suggests that some dimers are distributed in $L_1$ and $L_3$ as well.

\section{\label{sec:CONCLUSION}CONCLUSION}
We successfully synthesized the delithiated phase \dlro , the D-phase, with structures distinct from the stoichiometric \lro , the S-phase. For the first time, we identify the magnetic properties of the D-phase. We found that the Ru ions also form dimers as in the S-phase. There should be two kinds of dimers in the D-phase: \rufo --\rufo\ dimers as in the S-phase and additional \rufi --\rufi\ dimers. The latter should have a new molecular orbital configuration, in which no electrons occupy antibonding states. We find that above $T \approx 539$~K the D-phase exhibits a strong linear increase in the susceptibility. In analogy to the S-phase, this magnetic feature is most likely associated with the change from the dimer-solid to dimer-liquid states. Structural and magnetic properties indicate that the dimers in the dimer-solid state are located not only in the $L_2$ bonds but also in the $L_1$ and $L_3$ bonds. Such dimer distribution may lead to the observed superlattice structure.

\begin{acknowledgments}
We thank  G. Khaliullin, Z. Fisk, H. Takagi, M. Braden, and T. Fr\"{o}hlich for useful discussions.  We are grateful to Y. Honda for his support in the chemical composition analysis. We also acknowledge   J. Wright and M. Di Michiel for their supports.
This work was supported by the Grants-in-Aid for Scientific Research on Innovative Areas “Topological Materials Science” (KAKENHI~Grant No. 15H05852), as well as KAKENHI~Grant No. 26247060, from JSPS of Japan.
M.-P. J.-S. is supported by the Japanese government (MEXT) scholarship. 
\end{acknowledgments}

\appendix
\section{\label{sec:decomposition}Reaction of Li$_{2-x}$RuO$_3$ at high temperatures}

In the XRD spectra shown in Fig.~\ref{fig:XRDlowhi}, the peak intensities of the \ruo\ ($P4_{2}/mnm$) peaks at 625~K are markedly  higher than those at 583~K. See for example the peak at $2\theta = 2.57$. We find that this enhancement starts at $\approx 609$~K (not shown in the figure). The structure of the main phase after heating above $T \approx 609$~K  is equivalent to that  of the S-phase \lro . 

Besides, as shown in Fig.~\ref{fig:MTlowhi} the temperature dependence of the susceptibility of \dlro\ $x \approx 0.73$ changes substantially before and after heating to 700 K. The behavior after heating to 700~K becomes similar to that of pristine \lro\ (compare purple and black lines in Fig.~\ref{fig:MTlowhi}). While repeating magnetization measurements, the obtained curves remain almost equivalent to the purple curve in Fig.~\ref{fig:MTlowhi}.

These results indicate that, when Li$_{2-x}$RuO$_3$ is heated above $T~ \approx 609$~K, it decomposes into Li$_{2}$RuO$_3$ and RuO$_2$ according to the following reaction:\\
\begin{equation} \label{eq:decomposition}
\ce{Li_{2-\textit{x}}RuO3 ->[\text{heat in}][\text{air or He}] $\left(\dfrac{2-x}{2}\right)$ Li2RuO3 + $\dfrac{x}{2}$ RuO2 + $\dfrac{x}{4}$ O2}.
\end{equation}
This decomposition reaction seems to be triggered by the loss of oxygen. 

The minor phase \dlro\ with $x \approx 1.1$ (SG $R\bar{3}$), decomposes at lower temperatures. The peak at $2 \theta = 1.75^\circ$ seen as a shoulder in Fig.~\ref{fig:XRDlowhi}, characteristic of this phase, disappears at 497~K. The magnetization exhibits a small change of slope at this temperature (Fig.~\ref{fig:MTlowhi}).

\section{\label{sec:mix}First-order transition of the S-phase \lro\ }

\begin{figure}
\includegraphics[width=8cm]{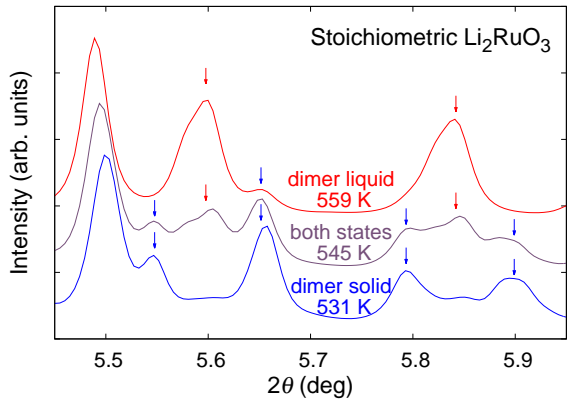} 
\caption{\label{fig:mix} (color online) High-energy XRD spectra of pristine \lro\ sample at three different temperatures: Below \td  $\sim 550$~K, close to \td , and above \td\ . It shows the coexistence of the dimer-solid and dimer-liquid states, which is typical of a first-order transition. The blue arrows indicate peaks characteristic of the dimer-solid state and red arrows peaks of the dimer-liquid state. The measurements with $E=87.5~\text{keV}$ ($\lambda =$ 0.14169~\AA) were performed at the beam line ID11 of ESRF.} 
\end{figure}

Initially the dimer transition of the pristine \lro\ was considered to be second order \cite{miura_new-type_2007}. However, recent studies indicate that the dimer transition is a first-order transition \cite{terasaki_ruthenium_2015, jimenez-segura_effect_2016}. Figure~\ref{fig:mix} shows the high-energy XRD spectra of pristine \lro\ at three different temperatures. At the temperature close to the dimer transition, the spectrum shows the combination of the structures of the dimer-solid and dimer-liquid states. This coexistence provides additional evidence for the first-order transition. 

\end{document}